\documentclass[11pt,onecolumn,floatfix,superscriptaddress,nofootinbib]{revtex4-1}

\usepackage{graphicx}                   
\usepackage{epsfig}
\usepackage{amsmath,amssymb}            
\usepackage{hyperref}                   
\usepackage{epstopdf}
\usepackage{placeins}
\usepackage{upgreek}
\usepackage{slashed}
\usepackage{color}
\usepackage{footmisc}
\usepackage{footnote}
\definecolor{orange}{rgb}{1,0.5,0}
\definecolor{brown}{rgb}{0.65, 0.16, 0.16}
\definecolor{phlox}{rgb}{0.87, 0.0, 1.0}

\graphicspath{{figs/}}                  

\begin{document}

\title{Generation of the CMB cosmic Birefringence through Axion-like particles, Sterile and Active neutrinos }
	
\author{S. Mahmoudi}
\email{s.mahmoudi@shirazu.ac.ir}
\affiliation{ Department of Physics, School of Science, Shiraz University, Shiraz 71454, Iran }
\author{ Mahdi Sadegh}
\email{m.sadegh@ipm.ir}
\affiliation{School of Particles and Accelerators, Institute for Research in Fundamental
	Sciences (IPM), P. O. Box 19395-5531, Tehran, Iran}
\author{Jafar Khodagholizadeh}
\email{gholizadeh@ipm.ir}
\affiliation{Farhangian University, P.O. Box 11876-13311, Tehran, Iran}
\author{Iman Motie}
\email{iman.motie@univ-tlse3.fr}
\affiliation{Universitè de Toulouse, UPS-OMP, IRAP, F-31400 Toulouse, France,\\
     CNRS, IRAP, 14 avenue Edouard Belin, F-31400 Toulouse, France }
\author{Alain Blanchard}
\email{alain.blanchard@irap.omp.eu }
\affiliation{Universitè de Toulouse, UPS-OMP, IRAP, F-31400 Toulouse, France,\\
     CNRS, IRAP, 14 avenue Edouard Belin, F-31400 Toulouse, France}
\author{She.-Sheng. Xue}
\email{xue@icra.it}
\affiliation{
ICRANet Piazzale della Repubblica, 10 -65122, Pescara, Italy,
\\
Physics Department, Sapienza University of Rome, 
Rome, Italy,\\
 INFN, Sezione di Perugia, 
Perugia, Italy,
\\
 ICTP-AP, University of Chinese Academy of Sciences, Beijing, China.
}

\begin{abstract}
 The cosmic birefringence (CB) angle refers to the rotation of the linear polarization plane of Cosmic Microwave Background (CMB) radiations when parity-violating theories are considered. We analyzed the Quantum Boltzmann equation for an ensemble of CMB photons interacting with the right-handed sterile neutrino dark matter (DM) and axion-like particles (ALPs) DM in the presence of the scalar metric perturbation. We used the birefringence angle of CMB to study those probable candidates of DM. It is shown that the CB angle contribution of sterile neutrino is much less that two other sources considered here. Next, we combined the results of the cosmic neutrinos' contribution and the contribution of the ALPs to producing the CMB birefringence and discussed the uncertainty on the parameter space of axions caused by the share of CMB-cosmic neutrino interaction in generating this effect. Finally, we plotted the EB power spectrum of the CMB and showed that this spectrum behaves differently in the presence of cosmic neutrinos and ALPs interactions in small $l$. Hence, future observed data for $C^{l}_{EB}$, will help us to distinguish the CB angle value due to the various sources of its production.
      
    \end{abstract}


    \maketitle

\section{\label{sec:level1}Introduction}

The detection of CMB radiation has been a major achievement of physics in the $20^{th}$ century, leading to the establishment of the standard cosmological model  ($\Lambda$CDM)
\cite{ Planck:2018vyg, SPT-3G:2021eoc, Komatsu:2014ioa, SPIDER:2021ncy, BICEP:2021xfz}. Nevertheless, recent and more precise measurements of the CMB photons along with other data from the $\Lambda$CDM model, propose the existence of new physics beyond the standard model (SM)\cite{Abdalla:2022yfr}. In this regard, experimental data related to the CMB birefringence are important observations confirming this phenomenon \cite{Komatsu:2022nvu}.\par

The CB effect indicates the rotation of the linear polarization plane of the CMB photons while traveling through the cosmos \cite{Carroll:1989vb, Harari:1992ea}. This rotation modifies the gradient and
curl of the polarization pattern of the CMB called E-mode and B-mode, respectively  \cite{Zaldarriaga:1996xe},
creating B-modes from E-modes. Since the $E$ and $B$ modes of the CMB transform differently under inversion of spatial coordinates, $E_{\ell m} \to (-1)^{\ell} E_{\ell m}$ and $B_{\ell m} \to (-1)^{\ell+1}B_{\ell m}$, the auto power spectra of these quantities are invariant under the parity transformation while the cross power spectrum changes sign. Hence, the existence of the non-zero values of $C_{\ell}^{EB}$ could be evidence for the parity violation phenomena such as axion-like pseudo-scalar fields, known as ALPs, coupling to photons \cite{Nakatsuka:2022epj, Murai:2022zur}.\par

Several studies have recently reported a non-zero value of CB using the latest CMB
polarization data \cite{Minami:2020odp,Diego-Palazuelos:2022dsq,PhysRevLett.128.091302}. The latest reported value for the rotation angle of the
CMB polarization is \cite{Minami:2020odp,PhysRevLett.128.091302} 
\begin{equation}
	\beta=0.342^{+0.094}_{-0.091}~\mathrm{deg}.
\end{equation}
Forthcoming and ongoing CMB experiments such as LiteBIRD \cite{LiteBIRD:2022cnt}, Simons Observatory (SO) \cite{SimonsObservatory:2018koc}, and CMB-S4 (S4) \cite{CMB-S4:2020lpa}.
are designed to make more sensitive polarization measurements, enabling us to improve current constraints on CMB birefringence angle ($\beta$).

One plausible explanation of the isotropic CMB birefringence is the interaction of photons with ALPs which start to move during or after the recombination epoch \cite{Abbott:1982af}. However, there might be other parity-violating phenomena that give rise to the CB effect. Indeed, CB can be considered a probe of new physics that breaks Lorentz and CPT symmetry \cite{Colladay:1998fq, Kostelecky:2002hh, Kostelecky:2007zz}. For instance, authors in \cite{Geng:2007va} proposed a CPT-even six-dimensional effective Lagrangian and considered the effect of neutrino density asymmetry on CB angle. Moreover, the weak interaction of the CMB and cosmic neutrino background (C$\nu$B) can produce non-vanishing EB power spectra at the level of one loop forward scattering in the presence of scalar perturbation, which generates part of the CB angle \cite{Mohammadi:2021xoh}. In addition, the impact of dipolar dark matter (DM), as well as sterile neutrino DM interaction with photons in producing the CB effect of the CMB has been studied by authors in \cite{Khodagholizadeh:2023aft}. In this paper, we introduce two new sources of CB, i.e., right-handed sterile neutrino DM and sterile neutrino QCD axion DM.\par

Over the past several decades, evidence has been generated by astrophysical and cosmological observations that cannot be explained unless DM exists besides normal matter composed of elementary particles in the SM of particle physics \cite{Green:2021jrr, Roszkowski:2004jc}. DM makes up nearly $85\%$ of the total matter in the universe \cite{Planck:2018vyg}. This observational evidence provides physically compelling reasons for new fundamental particles and interactions beyond the SM.  In this regard, a wide range of theoretically well-motivated DM candidates have been proposed including the weakly-interacting massive particle (WIMP) \cite{Jungman:1995df, Donato:2008jk, Roszkowski:2017nbc}, ALPs \cite{Peccei:1977hh} and
the sterile neutrino DM \cite{Abazajian:2012ys, Boyarsky:2018tvu}. \par

One of the most popular candidates for DM is sterile neutrinos with a mass of the order of a few keV confirmed to have all the criteria to play the role of
DM \cite{Abazajian:2001nj, Abazajian:2017tcc, Drewes:2016upu},  constituting all or a part of galactic DM halo. They can also satisfy the bounds from structure formation and the free streaming length of DM at early epochs \cite{Dodelson:1993je}. Moreover, their existence could be motivated by reasons of symmetry and by demand for giving mass to the SM active neutrinos. In this regard, various effective models have been provided enabling us to predict the existence of these particles \cite{Schechter:1980gr, Mohapatra:1979ia, Cheng:1980qt, Mohapatra:1980yp, Kusenko:2010ik, Cox:2017rgn}. Here, we are interested in studying the sterile neutrino particle, predicted by an effective model based on the fundamental symmetries and particle content of the SM \cite{Xue:2020cnw}. In fact, this scenario, which was motivated by the parity symmetry reconstruction at high energies without any extra gauge bosons, introduces three massive sterile neutrinos and the SM gauge symmetric four-fermion interactions giving rise to new effective interactions between sterile neutrinos and SM gauge bosons. We will investigate the interaction effects of this sort of sterile neutrino particle with photons on the rotation of the linear polarization plane of the CMB. \par

Another favored theoretical candidate for DM is the ALPs, being the generalizations of axions, which were originally motivated by solving the strong CP problem
through promoting the CP-violating phase $\theta$ to $a/f_{a}$ with $f_a$
known as the Peccei-Quinn (PQ) scale \cite{Peccei:1977hh, Weinberg:1977ma}. These particles emerge naturally from theoretical models of physics at high energies, including string theory, grand unified theories, and models with extra dimensions \cite{Irastorza:2018dyq, Svrcek:2006yi}. In the continuation of this path, the author in \cite{Xue:2020cnw} showed that the spontaneous breaking of global $U(1)$ chiral symmetry in the sterile right-handed neutrino sector results in generating mass for the sterile neutrinos, accompanied by a pseudoscalar Goldstone boson of PQ type axion $a$, known as sterile neutrino QCD axion, and a massive scalar boson of mass in the order of $10^2$ GeV. Notably, due to the very long lifetimes and tiny couplings to SM particles, they can be potential DM candidates. In one part of this paper, we consider sterile neutrino QCD axion as a candidate of DM and try to study this model through its interaction effect with photons on the CMB birefringence. Moreover, by taking into account the contribution of active neutrinos on producing CB angle, we discuss the constraint placed on the parameter space of the axion particles.\par

The rest of this paper is organized as follows:  Section \ref{CB} presents a general discussion about the relativistic Boltzmann equations as well as birefringence theory and briefly reviews how the CB angle affects CMB power spectra. We introduce the right-handed sterile neutrinos whose coupling to the photon would induce a rotation of the polarization plane of the CMB in Section \ref{CB1} and investigate how this candidate of DM may influence the CMB birefringence. Next, we provide a short review of the generation of the CB angle due to the C$\nu$B with the CMB following Ref. \cite{Mohammadi:2021xoh}. Section \ref{ALPs} is devoted to studying the CB effect due to the interaction between the CMB photons and ALPs. To this end, we first introduce the sterile neutrino QCD axion model and then discuss its interaction effect with the photon on the CMB birefringence in Section \ref{CB1}. In Section \ref{combination}, we combine the results of CB angle calculations due to the interaction of the CMB photons with cosmic neutrinos as well as ALPs and discuss the constraint on the axion's parameter phase space. Section \ref{cross} shows the EB power spectrum of the CMB in the presence of Thomson scattering and CMB-C$\nu$B and CMB-Axion. A summary and remarks are provided in section \ref{conclusion}.

\section{General discussion about Boltzmann equations and the CB theory}\label{CB}

The polarization properties of electromagnetic radiation can be described by Stokes parameters, which, for a propagating wave in the $\hat{z}$ direction are defined as
\begin{eqnarray}
	I &\equiv& \langle E_x^2 \rangle + \langle E_y^2 \rangle \,\,\,\,\,\,\,\,\,\,\,\,\,\,\,\,\,\,\,\,\,\,\,\,\,\,\,\,\,\,\,\,\,\,\,\,\,\,\,\,\,\,\,\,\,\,\,\,\,\,\,\,\,
	Q \equiv \langle E_x^2 \rangle - \langle E_y^2 \rangle \, , \nonumber \\
	U &\equiv& \langle 2E_xE_y \cos (\phi_x - \phi_y )\rangle \,\,\,\,\,\,\,\,\,\,\,\,\,\,\,\,\,\,\,\,\
	V\equiv \langle 2E_xE_y \sin (\phi_x - \phi_y )\rangle \, .
	\label{eqn:stokes}
\end{eqnarray}
where $I$ indicates the total intensity of radiation, $U$ and $Q$ describe the linear polarization of photons and parameter $V$ represents the net circular polarization or the difference between left- and right-circular polarization intensities. Moreover, the amplitudes and phases of waves in the $x$ and $y$ directions are given by ($E_x$,$\phi_x$) and ($E_y$,$\phi_y$), respectively, and $\langle\cdot\cdot\cdot\rangle$ represents time averaging. Note that the parameters $Q$ and $U$ are influenced by the orientation of the coordinate system while $V$ and $I$ are coordinate-independent. While the Stokes parameters $Q$ and $U$ offer a local definition of polarization, their coordinate dependence makes them rather inconvenient for cosmological interpretation. As a solution to this problem, a set of linear combinations of polarization parameters $Q$ and $U$ can be introduced as $\Delta^{\pm}_{\text{P}}({\bf k})=(Q\pm i U)({\bf k})$, which are the reference frame-independent parameters and $P$ stands for polarization. Moreover, to study the polarization characteristics of the CMB photons in the context of cosmology, it is common to separate the two polarization quantities $\Delta^{\pm}_{\text{P}}({\bf k})$ into a curl-free part (E mode) and a divergence-free part (B mode) as follows:
\begin{eqnarray}
	\Delta_{\text{E}}({\bf k})&\equiv&-\dfrac{1}{2}\left[\bar{\eth}^{2}\,\Delta^{+}_{\text{P}}({\bf k})+\,\eth^{2}\Delta^{-}_{\text{P}}({\bf k}) \right],\label{BE11}\\
	\Delta_{\text{B}}({\bf k})&\equiv&\dfrac{i}{2}\left[\bar{\eth}^{2}\,\Delta^{+}_{\text{P}}({\bf k})-\eth^{2}\Delta^{-}_{\text{P}}({\bf k})\right],\label{BE12}
\end{eqnarray}
where $\eth $ and $\bar{\eth}$ are spin-raising and lowering operators, respectively \cite{Zaldarriaga:1996xe}. These modes provide an alternative description
of CMB linear polarization which, unlike $Q$- and
$U$-modes, have the advantage of being rotationally invariant like the temperature
and no ambiguities linked to the rotation of the coordinate system arise.\par

In the framework of quantum mechanics, the polarization of an ensemble of photons can be explained by the following density operator:
\begin{equation}\label{stoke7}
	\hat{\rho}=\frac{1}{tr(\hat{\rho})}\int\frac{d^3p}{(2\pi)^3}\rho_{ij}({\bf k})\hat{D_{ij}}({\bf k}),
\end{equation}
where $\rho_{ij}$ shows the density matrix components in the phase space, ${\bf k}$ represents the momentum of cosmic photons and $\hat{D_{ij}}({\bf k})=a_i^\dagger({\bf k})a_j({\bf k})$ is the number operator of photons.
Indeed, the density matrix encodes the
information of intensity and the polarization of the photon ensemble and its dependency on the Stokes parameters is given by 
\begin{eqnarray}\label{eq:rho}
	\hat{\rho}_{ij}\equiv\frac{1}{2}\left(
	\begin{matrix}
		I+Q & U-iV \\
		U+iV&  I-Q \\
	\end{matrix}
	\right).
\end{eqnarray}

The time evolution of the density matrix components $\rho_{ij}({\bf{k}})$ is given by the quantum Boltzmann equation as \cite{Kosowsky:1994cy}
\begin{eqnarray}\label{forward}
	(2\pi)^3\delta^3(0)(2k^0)\dfrac{d}{dt}\rho_{ij}({\bf{k}})=i\langle[H_{I}^0(t);D_{ij}^{0}({\bf{k}})]\rangle -\dfrac{1}{2}\int dt \langle[H_{I}^0(t);[H_{I}^0(t);D_{ij}^{0}({\bf{k}})]]\rangle,
\end{eqnarray}
where $H_{I}^0(t)$ is the first order of the interaction Hamiltonian
\begin{equation}\label{HI}
	H_{I}^0(t)=-\frac{i}{2}\int\limits_{-\infty}^{\infty}\,dt'T\{H(t),H(t')\},
\end{equation}
where $T$ represents a time-ordered product and $H$ denotes the interaction Hamiltonian, relating to the interaction Hamiltonian density $\mathcal{H}(x)$ as follows
\begin{equation}
	H(t)=\int d^{3}{\bf{x}}\,\mathcal{H}(x).
\end{equation}
\color{black}
Notably, the first term on the right-hand side of Eq. \eqref{forward} is called the forward scattering term, which is proportional to the amplitude of the scattering, whereas the second one is known as the higher-order collision term, giving the scattering cross-section, which is highly sub-dominant compared to the first term.\par

In the framework of the SM of cosmology, due to the parity symmetry, $E$ and $B$ mode polarizations are not correlated. Indeed, 
since the E-modes are scalars under parity but B-modes are pseudo-scalar, the two modes must be uncorrelated in the absence of parity-violating physics. However, the presence of some parity-violating as well as Lorentz symmetry-breaking interactions cause a difference in the phase velocities of the right- and left-hand helicity states of photons and subsequently the rotation of the plane of linear polarization in the sky by an angle $\beta$ 
\begin{equation}
	\label{eqn:rotation}
	\Delta^{\pm}_{\text{P}}\mapsto\Delta^{\pm}_{\text{P}}e^{\pm2i\beta},
\end{equation}
where the function $\beta$, called \emph{birefringence angle}, measures the amplitude of deviation from the SM. Hence, a part of $E$-modes' polarization transfers into $B$-mode ones, resulting in the non-zero $EB$ mode correlation. 
Such an effect could have left measurable imprints in the CMB angular power spectra $C_{\ell}$'s. Indeed, the presence of parity violation interaction induces a rotation of the CMB angular power spectra as follows\cite{Minami:2019ruj}:
\begin{eqnarray}
	C^{EE,o}_{\ell}&=&\cos^2(2\beta){C}^{EE}_{\ell}+\sin^2(2\beta){C}^{BB}_{\ell},\\
	C^{BB,o}_{\ell}&=&\cos^2(2\beta){C}^{BB}_{\ell}+\sin^2(2\beta){C}^{EE}_{\ell},\\
	C^{EB,o}_{\ell}&=&\dfrac{1}{2}\sin(4\beta)({C}^{EE}_{\ell}-{C}^{BB}_{\ell}),\label{ClEB}
\end{eqnarray} 
where ${C}^{EE}_{\ell}$ is the $E$-mode power spectra in the epoch of the recombination, and the superscript "o" denotes
the observed value. Moreover, $\beta={0.342^\circ} ^{+0.094^\circ}_{-0.091^\circ}$ ($68\%\text{C.L.}$) is the value of the CB angle reported by using Planck data release \cite{Eskilt:2022cff}. Note that $C^{EB,o}_{\ell}$ has become
the most sensitive probe
of parity violation in the current era of CMB experiments
with low polarization noise \cite{Komatsu:2022nvu}.\par

\section{CB angle from sterile and cosmic SM neutrinos interacting with CMB}\label{CB1}

In this section, we first study right-handed sterile neutrinos (according to the new scenario \cite{Xue:2020cnw}) whose coupling to the photon would induce a rotation of
the polarization plane of the CMB, and try to investigate their properties through CMB birefringence. Next, we re-analyze some results of \cite{Mohammadi:2021xoh}, regarding the contribution of C$\nu$B-CMB interaction in producing the CB angle. \\
Here, we briefly describe the ultraviolet (UV) completion of the low-energy effective model adopted in this study. On the one hand, as shown in low-energy experiments, the SM possesses parity-violating (chiral) gauge symmetries $SU_c(3)\times SU_L(2)\times U_Y(1)$. 
On the other hand, as a well-defined QFT, the SM should regularize at the high-energy cutoff $\Lambda_{\rm cut}$, fully preserving the SM gauge symmetries.  
A natural UV regularization is provided by a theory of new physics at $\Lambda_{\rm cut}$, for instance, quantum gravity. However, the theoretical inconsistency between the SM bilinear Lagrangian and the natural UV regularization, due to the No-Go theorem~\cite{NIELSEN1981173,NIELSEN1981219}, which implies right-handed neutrinos and their quadrilinear four-fermion operators. We adopt the four-fermion operators of the torsion-free Einstein-Cartan Lagrangian with SM leptons $\psi^{f}$ and three right-handed sterile neutrinos $\nu^{f}_{_R}$~\cite{Xue:2016dpl, Xue:2016txt}:
\begin{eqnarray}
	{\mathcal L}
	&\supset &-G_{\rm cut}\sum_{f=1,2,3}\left(\, \bar\nu^{fc}_{_R}\nu^{f}_{_R}\bar\nu^{f}_{_R} \nu^{fc}_{_R}
	+\, \bar\nu^{fc}_{_R}\psi^{f}_{_R}\bar\psi^{f}_{_R} \nu^{fc}_{_R}\right)+{\rm h.c.},
	\label{art1}
\end{eqnarray}
where $G_{\rm cut}\propto \Lambda^2_{\rm cut}$ and the two-component Weyl fermions $\nu^{f}_{_R}$ and $\psi^{f}_{_R}$ 
zre respectively, the eigenstates of the SM gauge symmetries. Three right-handed sterile neutrinos are possible candidates for DM particles. 
Their interactions with SM leptons $\psi^f_R$ effectively induce in low energies the one-particle-irreducible (1PI) vertexes interacting with the SM gauge bosons \cite{Xue:2020cnw}
\begin{eqnarray}
	\mathcal{L}&\supset & \mathcal{G}^{W}_{R}(g_{w}/\sqrt{2})\big[(U^{\ell}_{R})^{\dagger}U^{\nu}_{R}\big]^{\ell\ell'}\bar\ell_R\gamma^{\mu}N^{\ell'}_RW^{-}_{\mu} 
	+{\mathcal{G}^Z_R}~({g_w}/{\sqrt{2}})\bar \nu^\ell_R\gamma^\mu\nu^\ell_R Z^0_\mu
	+\mathcal{G}^\gamma_R(e)\bar\nu^\ell_R\gamma^\mu\nu^\ell_R A_\mu + {\rm h.c.}\,
	\label{Lagrangian}
\end{eqnarray}
The effective couplings ${\mathcal{G}_R^{W,Z,\gamma}}$ represent in the momentum space the 
1PI interacting vertexes.
The $SU_L(2)$ gauge coupling $g_w=e/\sin\theta_W$ is defined by the electric charge $e$ and Weinberg angle $\theta_W$ and relates to 
the Fermi constant $G_{F}/\sqrt{2}=g_{w}^{2}/8M_{W}^{2}$ and 
the $W^\pm$ gauge boson mass $M_{W}$.

The right-handed neutrino $\nu^\ell_R$ is in the same family of the charged lepton $\bar\ell_R$. The right-handed doublets $(\nu^\ell_R, \ell_R)$ are gauge eigenstates in gauge interacting basis, and $(N^\ell_R, \ell_R)$ are the corresponding mass eigenstates in the mass basis. $U^{\nu}_{R}$ and $U^{\ell}_{R}$ are unitary matrices $3\times 3$ ($\ell=e,\mu,\tau$). Note that the mixing matrix 
$V_{\ell N}\equiv \big[(U^{\ell}_{R})^{\dagger}U^{\nu}_{R}\big]$ is not the PMNS one  $\big[(U^{\ell}_{L})^{\dagger}U^{\nu}_{L}\big]$.  
Summation over three lepton families is performed in Eq.~(\ref{Lagrangian}) and no flavor-changing-neutral-current (FCNC) interactions occur. The effective operator $\mathcal{G}^{W}_{R}$ contributes to vector boson fusion (VBF) processes (see the left Feynman diagram in Figure 1 of 
Ref.~\cite{Collaboration2022}).
The sterile neutrinos' mixing $|V_{\ell N}|$ and masses $M_{N^\ell}$ are constrained \cite{Sirunyan2018,Sirunyan2019,Collaboration2022}. 
Here, we neglect mixing by approximating $V_{\ell N}\approx 1$, and consider only the first lepton family $(N^e_R, e_R)$. 

Moreover, for simplicity, ${\mathcal{G}_R^{W, Z,\gamma}}\approx {\mathcal{G}_R}\ll 1$ is assumed. The ${\mathcal{G}_R}$ is the unique parameter constrained by experiments and observations.
The upper limit of ${\mathcal{G}_R}$ value should be smaller than $10^{-4}$. It is constrained by the $W^\pm$ decay width \cite{Haghighat:2019rht}, studying double beta-decay $0\nu\beta\beta$ experiment \cite{Pacioselli:2020yzx}, $W$ 
boson mass tension \cite{Xue:2022mde}, the precision measurement of fine-structure constant $\alpha$ \cite{Xue:2020cnw} and regarding $N^e_R$ as a DM particle in the XENON1T experiment and astrophysical observations \cite{Xue:2020cnw}.\par

It is worth discussing the constraint on ${\mathcal{G}_R}$ which comes from the possibility of $N^e_R$ being a DM in more detail. Indeed, the introduced right-handed sterile neutrinos, depending on the value of the mass and ${\mathcal{G}_R}$ parameter, can be fit to a warm DM scenario. Nevertheless, the right-handed neutrinos can decay
radiatively into the SM neutrinos as $N_R^l\rightarrow \nu_L^l + \gamma$ with the following decay width 
\begin{align}
	\label{decayr}
	\Gamma(N_R^l\rightarrow \nu_L^l + \gamma)\approx
	\left( \frac{\alpha \,g_{w}^4}{1024\,\pi^4}\right)m_l^2(m^l_{N})^3\mathcal{G}_R^2,
\end{align}
where $\alpha$ is the fine-structure constant, and for $N^e_R$ at low energy scale, we have:
\begin{eqnarray}\label{NRe}
	\Gamma(N_R^e\rightarrow \nu_L^e + \gamma)&=&1.57\times 10^{-19}s^{-1} \left(\frac{\mathcal{G}_R}{10^{-4}} \right)^{2}\left(\frac{m_{e}}{511 \,{\text{KeV}}} \right)^{2}\left( \frac{M_{N}^{e}}{100\, {\text{KeV}}}\right)^{3}.
\end{eqnarray}
The assumption of right-handed neutrinos being DM particles requires that their lifetimes be longer than the age of the Universe  $t_\text{\tiny{Universe}}=4.4\times10^{17}\,\text{sec}$, leading to the following constraint on ${\mathcal{G}_R}$:
\begin{eqnarray}
	\mathcal{G}_{R}\lesssim 3.8 \times 10^{-4}\left(\frac{100~ {{\rm KeV}}}{m^e_{N}}\right)^{3/2}.
	\label{gRcon}
\end{eqnarray}
It indicates that sterile neutrinos weakly interact with SM gauge bosons, compared with active neutrinos. However, as a candidate of warm dark matter particles, sterile neutrinos may have a larger cosmic abundance than active neutrinos. It is deserved to study sterile neutrinos' cosmological effects.

In this article, using sterile neutrinos interacting with SM particles (\ref{Lagrangian}), we calculate sterile neutrino contributions to the CB angle, comparing with active neutrino contributions \cite{Khodagholizadeh:2023aft}. Details of calculations are presented in Appendix \ref{appendix}. As a result, we combine eq.(\ref{gRcon}) with results of cross-correlation power spectra (eq.(\ref{tauN}))
to obtain the constraint on sterile neutrino contributions to the CMB CB angle,
\color{black}

\begin{equation}
	\beta =\frac{1}{2}\bar\tau_{\text{\tiny{N}}}\leq 1.5\times 10^{-3}\,\mathcal{G}_{R}^{2}\,\,rad\,\,\left(\frac{KeV}{M_{\text{\tiny{N}}}}\right),
	\label{sresult}
\end{equation} 
where $\bar\tau_{\text{\tiny{N}}}$ is the maximum value of the effective sterile neutrino opacity (\ref{tauN1}) near the last scattering,  
\begin{equation}
	\tilde{\tau}_{\text{\tiny{N}}}\approx 3\times 10^{-3}\,\mathcal{G}_{R}^{2}\,\left(\frac{KeV}{M_{\text{\tiny{N}}}}\right)\,
	\bigg(\frac{\rho _{\text{\tiny{N}}}^{0}}{10^{-47}\,GeV^{4}}\bigg)\,\,\bigg(\frac{z'}{10^3}\bigg).
	\label{tauN1}
\end{equation}
obtained in Appendix \ref{appendix}. Thus, we conclude that a sterile neutrino DM with a mass of around $100KeV$ can contribute to producing CB angle up to $\beta\approx 1.5\times 10^{-9} rad$ at most.

It is important to compare the sterile neutrino contribution (\ref{sresult}) and (\ref{tauN1}) with active neutrino one \cite{Mohammadi:2021xoh}. Therefore, we re-analyze Eq.~(13) of Ref.~\cite{Mohammadi:2021xoh} to estimate the maximum value of $\bar \tau_{\nu}$ near the last scattering surface for C$\nu$B-CMB interaction:
\begin{eqnarray}
	&&\tilde\tau_{\nu}= \tau_{\nu}(z_l)\simeq 3.43573\delta_\nu (\frac{n_{\nu}^0}{340cm^3})~rad. \label{kappabar1}
\end{eqnarray}
where $\delta_l$ is the relative density asymmetry of Dirac cosmic neutrinos and anti-neutrinos. Employing the fact that 
\begin{equation}
	\beta^{\nu}=\frac{1}{2}\tilde\tau_{\nu}\simeq 1.717\,\,\delta_\nu rad, \label{betanu}
\end{equation}
the above estimation results in an upper bound over $\delta_\nu<3\times 10^{-3}$ to consistent with $\beta=0.342^{+0.094}_{-0.091}~\mathrm{deg}$ ($68\%\text{C.L.}$)\cite{Eskilt:2022cff,Minami:2020odp}.

As can be seen, the contribution of CB angle due to the sterile neutrino is too much small than C$\nu$B. So in comparison with other sources of generating CB angle that are considered here, can be negligible. 

\section{CB angle from ALPs and CMB interactions} \label{ALPs}

ALPs are generalizations of QCD axion which is a well-motivated solution to the strong CP problem \cite{Peccei:1977hh,Peccei:1977ur}. These particles, are generically predicted
in string theory, are understood as pseudo-scalar fields and interact with the SM through similar couplings to the QCD axion. However, they populate a much broader region of mass-coupling parameter space compared to the QCD axion. Due to this feature, they are considered a natural cold DM candidate \cite{Ferreira:2018wup,Sandvik:2002jz}.\par

Here, we study a particular ALP induced by the sterile-neutrino QCD axion model \cite{Xue:2020cnw}. As we will show, this particle can contribute to generating the CB angle of the CMB. Besides, we will employ this effect to constrain the mass-coupling parameter space of the model.


In the theoretical scenario for sterile neutrinos and interactions (see Sec.~\ref{CB1}) we discuss the sterile-neutrino QCD axion. Analogously to the top-quark mass generated in the $t\bar t$ condensate model $\bar t_{_L}t_{_R}\bar t_{_R} t_{_L}$ \cite{Bardeen1990}, the heaviest sterile neutrino Majorana mass is generated by the self-interaction (\ref{art1}) 
by spontaneous Peccei-Quinn (PQ) global symmetry breaking at the electroweak scale $v\approx 246$ GeV \cite{Xue2015, Xue:2016dpl} \footnote{The other two sterile neutrinos' Majorana masses are attributed to explicit symmetry breaking induced by (\ref{Lagrangian}).}.
The corresponding Goldstone mode, which is a pseudo scalar bound state of sterile neutrino and anti-neutrino $(\bar\nu^{c}_{_R}\gamma_5\nu_{_R})$, plays the role of the Peccei-Quinn QCD axion with decay constant $f_a\approx v$. Such a sterile-neutrino QCD axion $\phi$ coupling to two photons \cite{Xue:2020cnw}
\begin{eqnarray}
	\mathcal{L}_{eff}\supset g^s_{a\gamma}\frac{1}{f_a}\frac{e^2}{32\pi^2}\phi F_{\mu\nu}\tilde{F}^{\mu\nu},
	\label{sagg}
\end{eqnarray}
where $g_{a\gamma}^s$ is the coupling constant, 
$F^{\mu\nu}$ is the electromagnetic tensor, 
$\tilde{F}^{\mu\nu}\equiv\frac{1}{2}\epsilon^{\mu\nu\rho\sigma}
F_{\rho\sigma}$ is its dual and the $\epsilon^{\mu\nu\rho\sigma}$ indicates the Levi-Civita antisymmetric tensor which guarantees the CP violation.
Moreover, it is notable that
$g_{a\gamma}^s\approx (\mathcal{G}_R)^2\ll 1$, differently from $g_{a\gamma}\sim {\mathcal O}(1)$ in Peccei-Quinn QCD axion case. 
The reason is that the 1PI operator (\ref{sagg}) in low energies is effectively induced from the sterile neutrinos and SM gauge bosons interactions (\ref{Lagrangian}). 
The relationship between axion mass $m_a$ and axion-photon coupling $g^{s}_{a\gamma}$ for the sterile QCD axion model is
\begin{eqnarray}
	m_a ({\rm eV}) &=& 2.69\times 10^{5} \sum_q \ln\left(\frac{m^M_3}{m_{q}}\right) \mathfrak{g}^{s}_{a\gamma}({\rm GeV}^{-1}),
	\label{mg} 
\end{eqnarray}
where $\mathfrak{g}^{s}_{a\gamma}$ is defined by $\mathfrak{g}^{s}_{a\gamma}=\frac{ g^{s}_{a\gamma}}{f_a}\frac{e^2}{8\pi^2}\approx \frac{\mathcal{G}_R^2}{f_a}\frac{e^2}{8\pi^2}$.  
To compare and contrast, we plot in Figs.~\ref{param1} and \ref{param} the results of other QCD axion models and the sterile QCD axion relation (\ref{mg}) 
for $\sum_q\ln (m^M_3/m_{q})\in [1-10]$.  
The 
sterile neutrino QCD axion is a candidate for superlight DM particles.

\subsection{CB angle due to sterile-neutrino QCD axion and CMB photon interaction}

Working in natural units $c= \hbar =k_B=1$ and Lorentz-Heaviside units $\epsilon_0=\mu_0=1$,
the parts of the Lagrangian depending on the ALPs and photon fields can be written as
\begin{equation}\label{eq:L_a}
	{\cal L}_{a\gamma}=-\frac{1}{4}F_{\mu\nu}F^{\mu\nu}+\frac{1}{2}\partial_\mu \phi \partial^\mu \phi+
	\frac{1}{4}\mathfrak{g}^{s}_{a\gamma}\,\phi\,F_{\mu\nu}\tilde F^{\mu\nu}-V_\phi(\phi)\,,
\end{equation}
where $V_\phi(\phi)$ is the
effective axion potential which can be expanded as $V_\phi(\phi)=\frac{1}{2}m_\phi^2\phi^2+{\cal O}(\phi^3)$ around $\phi=0$,
with $m_\phi$ the effective axion mass.\par

Utilizing the Euler-Lagrange equations leads to the following equations of motion 
\begin{eqnarray}
	\label{eq:phi}
	\Box \phi\equiv\nabla_{\mu} \nabla^{\mu} \phi&=&-m_\phi^2 \phi+ \frac{\mathfrak{g}^{s}_{a\gamma}}{4} F_{\mu\nu} \tilde{F}^{\mu\nu}\,,\\
	\label{eq_F}
	\nabla_{\mu}F^{\mu\nu}&=&-\mathfrak{g}^{s}_{a\gamma}(\nabla_{\mu}\phi)\tilde{F}^{\mu\nu}\,,\\
	\nabla_{\mu}\tilde{F}^{\mu\nu}&=&0\;.
\end{eqnarray}
Considering the equation of motion of the photon field and employing the definition of the electromagnetic tensor $F^{\mu\nu}\equiv\nabla^{\mu}A^{\nu}-\nabla^{\nu}A^{\mu}$,  
Eq.~(\ref{eq_F}) reads as
\begin{equation}
	\label{eq_F2}
	\Box A_{\nu}-\nabla_{\nu}\left(\nabla_{\mu}A^{\mu}\right)-{R^{\mu}}_{\nu} A_{\mu}=-\frac{\mathfrak{g}^{s}_{a\gamma}}{2} (\nabla_{\mu} \phi) {{\epsilon^{\mu}}_{\nu}}^{\rho\sigma}F_{\rho\sigma}\;.
\end{equation}
In Lorenz gauge (generalized to curved spacetime), $g^{\alpha\beta}\nabla_\alpha A_\beta = 0$, the field equations become
\begin{align}
	\Box A_\nu - {R^{\mu}}_\nu A_\mu &= - g_{\nu \alpha}  (\mathfrak{g}^{s}_{a\gamma} \partial_{\mu}\phi ) \epsilon^{\mu\alpha\lambda\rho} \partial_\lambda A_\rho,
	\label{eq:AEoM}
\end{align}
where we have used $\epsilon^{\mu\alpha\lambda\rho} \nabla_\lambda A_\rho = \epsilon^{\mu\alpha\lambda\rho} \partial_\lambda A_\rho$ owing to the symmetries of the Levi-Civita tensor and Christoffel symbols.

For a spatially flat Friedmann-Robertson-Walker universe, the metric is:
\begin{equation}
	ds^2=-dt^2+a^2(t) d \boldsymbol{x}^2=a^2(\eta)\left[-d\eta^2+d \boldsymbol{x}^2 \right]\,,
\end{equation}
where $t$ is the cosmic time, $\eta$ is the conformal time 
and $\boldsymbol{x}$ indicates the space coordinates.
As a specific case, we will consider $\phi = \phi(\eta)$, and investigate solutions to the photon equations of motion in Coulomb gauge ($\nabla\cdot\boldsymbol{A}=0$) by assuming the plane wave propagates along the $z$-axis. Under these assumptions, the two relevant components of Eq.~(\ref{eq_F2}) will be obtained as follows \cite{Fedderke:2019ajk}
\begin{eqnarray}
	A_{x}^{\prime\prime}\left(\eta,z\right)-\frac{\partial^2 A_{x}\left(\eta,z\right)}{\partial z^2}&=& \mathfrak{g}^{s}_{a\gamma} \phi^{\prime} \frac{\partial A_{y}\left(\eta,z\right) }{\partial z}\,,\\
	A_{y}^{\prime\prime}\left(\eta,z\right)-\frac{\partial^2 A_{y}\left(\eta,z\right)}{\partial z^2}&=&- \mathfrak{g}^{s}_{a\gamma} \phi^{\prime} \frac{\partial A_{x}\left(\eta,z\right) }{\partial z}\,.
\end{eqnarray}  
In the Fourier space, the above equations become:
\begin{eqnarray}
	\tilde{A}_{x}^{\prime\prime}(k,\eta)+k^2 \tilde{A}_{x}(k,\eta)+ \mathfrak{g}^{s}_{a\gamma} \phi^{\prime} i k \tilde{A}_{y}(k,\eta)=0\,,\label{xcom}\\
	\tilde{A}_{y}^{\prime\prime}(k,\eta)+k^2 \tilde{A}_{y}(k,\eta)- \mathfrak{g}^{s}_{a\gamma} \phi^{\prime} i k \tilde{A}_{x}(k,\eta)=0\,\label{ycom}.
\end{eqnarray}
where $k$ is the Fourier conjugate of $z$ and we have used the definition  $\tilde{A}_{x,y}(k,\eta)=(2\pi)^{-1}\int e^{i k z} A_{x,y}(\eta,z) dz$. 
By introducing the definite-helicity transverse field variables \cite{Komatsu:2022nvu}
\begin{equation}
	\tilde{A}_{\pm}(k,\eta)=\tilde{A}_{x}(k,\eta) \pm i \tilde{A}_{y}(k,\eta),
\end{equation}
equations (\ref{xcom}) and (\ref{ycom}) can be decoupled as
\begin{equation}
	\label{eq:A}
	\tilde{A}_{\pm}^{\prime\prime}(k,\eta)+\left[ k^2 \pm \mathfrak{g}^{s}_{a\gamma} \phi^{\prime}  k \right] \tilde{A}_{\pm}(k,\eta)=0\,.
\end{equation}  
Therefore, the equation of motion yields a helicity-dependent dispersion relation. When the effective angular frequency, $\omega_\pm^2\equiv k^2\mp k\mathfrak{g}^{s}_{a\gamma}\phi'$, varies slowly with time within one period, $|\omega_{\pm}'|/\omega_{\pm}^2\ll 1$, a WKB solution is $A_{\pm}\simeq
(2\omega_{\pm})^{-1/2}e^{-i\int d\eta\omega_{\pm}+i\delta_{\pm}}$, where $\delta_{\pm}$ is the initial phase of the $\pm$ states.
Furthermore, the phase velocity is given by
\begin{equation}
	\label{eq:phasevelocity}
	\frac{\omega_\pm}{k}\simeq 1\mp \frac{\mathfrak{g}^{s}_{a\gamma}\phi'}{2k}\,,
\end{equation}
when the second term is small. Indeed, the second term is very small: it is at most of the order of the ratio of the photon wavelength and the size of the visible Universe, $(k\eta)^{-1}$. However, the impact on CMB accumulates over a very long time (more than 13 billion years), which makes $\int d\eta(\omega_+-\omega_-)$ large enough to be observable. Therefore, we keep $\omega_\pm$ in the phase but set $\omega_{\pm}\simeq k$ in the amplitude of the WKB solution.\par

The difference in the phase velocity leads to the rotation of the plane of linear polarization\cite{carroll/field/jackiw:1990,carroll/field:1991,harari/sikivie:1992}.  In the CMB convention defined earlier,
the Stokes parameters for linear polarization are $Q\propto \cos[\int d\eta(\omega_+-\omega_-)-(\delta_+-\delta_-)]$ and $U\propto -\sin[\int d\eta(\omega_+-\omega_-)-(\delta_+-\delta_-)]$. This induces a modification of the Boltzmann equation for the evolution of polarization
perturbations \cite{Komatsu:2022nvu}
\begin{eqnarray}
	\dot{\Delta}^{(a)}_{P} + iK\mu \Delta^{(a)}_{P}
	=\dot\tau_{e\gamma}\Biggl[ -\Delta^{(a)}_{P}
	\left .\sum_{m}\sqrt{\frac{6\pi}{5}}
	\ _{\pm 2} Y_2^m
	S_{P}^{(m)} \right]\pm i 2\beta^{\prime} \Delta^{(a)}_{P},
	\label{fps}
\end{eqnarray}
where superscript $a$ stands for axion. $\Delta _{P}^{(a)}=Q^{(a)}\pm i U^{(a)}$, $S_P^{(a)}({\bf
	k},\eta) \equiv \Delta_{T2}^{(a)}({\bf k},\eta)+
12\sqrt{6}\Delta_{+,2}^{(a)}({\bf k},\eta)+12\sqrt{6}\Delta_{-,2}^
{(a)}({\bf k},\eta)$ is the source term of generating polarization, and $\beta^{\prime}=-(\omega_{+}-\omega_{-})/2$.
Hence, the birefringence angle is given by
\begin{equation}
	\label{eq:beta}
	\beta=\frac12\left(\delta_+-\delta_-\right)-\frac12\int d\eta\left(\omega_+-\omega_-\right)\,.
\end{equation}
Without loss of generality, we set $\delta_+-\delta_-=0$ from now on. Using equation~(\ref{eq:phasevelocity}), we find $\beta=\frac12 \mathfrak{g}^{s}_{a\gamma}\int_{\eta_{\mathrm{LS}}}^{\eta_0} d\eta\phi'$, where the subscripts `LS' and `0' denote the time of the last scattering and the present-day time, respectively. Therefore, the interaction between the sterile-neutrino QCD axion and CMB photon leads to  a rotation of the polarization plane and produces a birefringence angle as follows \cite{Fedderke:2019ajk, Komatsu:2022nvu}
\begin{equation}
	\beta = \frac{\mathfrak{g}^{s}_{a\gamma}}{2}\Delta\phi \equiv \frac{\mathfrak{g}^{s}_{a\gamma}}{2}(\phi_0 - \langle\phi_{\rm LSS}\rangle). \label{betaaxion}
\end{equation}

\begin{figure}[t]
	\centering
	\includegraphics[scale=0.5]{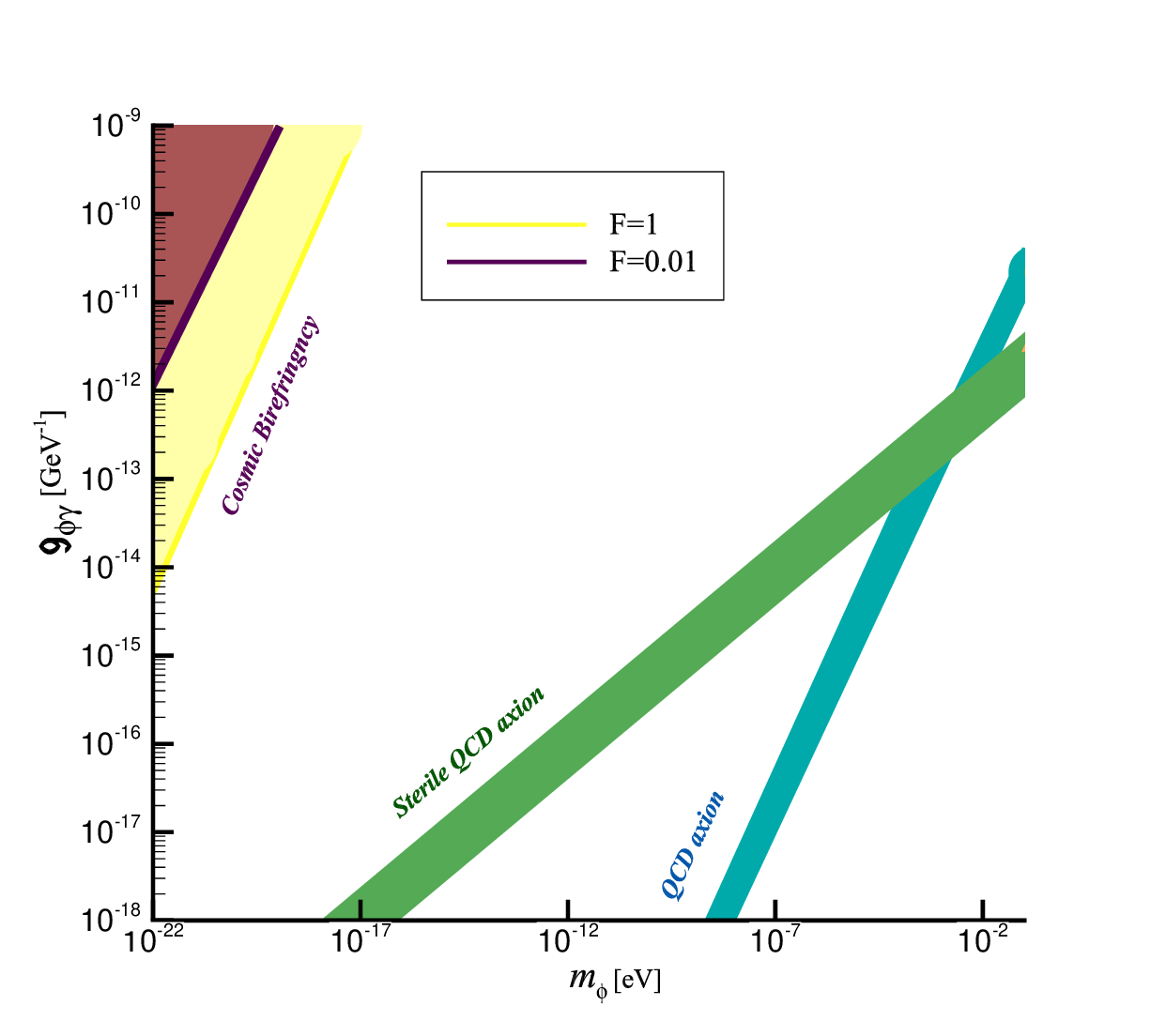}
	\caption{Parametric space of axion coupling ($g_{a\gamma}$)-mass ($m_a$). The yellow and purple shaded regions have been excluded by the CB effect of the CMB. The light green region is the permitted region for sterile QCD axion. } 
	\label{param1}
\end{figure}

In order to calculate the total change of $\phi$ from the last surface scattering until today, we employ the relation for axion energy density $\rho_a= (1/2)\,m_a^2\,\phi^2\,$ and 
\begin{eqnarray}
	\rho_a(z_{*}) &\simeq & \mathbb{F} \, \Omega_{\rm crit} \, (1+z_*)^3 \, \rho_{\rm crit,0},
\end{eqnarray} 
where $\rho_{\rm crit,0}$ is the present-day critical density and $z_*$ stands for the redshift of the last scattering surface. Moreover, $\mathbb{F} = \Omega_a / \Omega_{\rm c}$ indicates the ALP DM energy density fraction in which $\Omega_a = \rho_{a,0}/\rho_{\rm crit,0}$ is the present-day density parameter of ALP and $\Omega_{c}$ refers to the cold DM density parameter.
Making use of the local Galactic DM density $\rho_{a,\,{\rm local}} \simeq 0.3 \,{\rm GeV}\,{\rm cm}^{-3}$ and cosmological parameter values from Planck \cite{Planck:2018vyg}, and after subtracting the local ALP field value, we obtain
\begin{eqnarray}\label{eqphi}
	\Delta \phi \simeq
	1.5 \times{ 10^{-4}} \,\,\mathbb{F}^{1/2} {\rm GeV}\,\,\,\,\,\,
	{\left( \frac{\mu \text{eV}}{m_a} \right)}
	\!\!\left( \frac{\Omega_c}{0.264} \right)^{\!1/2}
	\left(\frac{\rho_{\rm crit,0}}{8.098\times 10^{-47} h^2 \,{\rm GeV}^4}\right)^{1/2}
	\!\left(\! \frac{z_*}{1089} \!\right)^{\!\!3/2}  
\end{eqnarray}
which, by using Eq. (\ref{betaaxion}), leads to the following relation
\begin{equation}
	\mathfrak{g}^{s}_{a\gamma}\simeq1.3 \times {10^{4}}\,\,\mathbb{F}^{-1/2}\beta\,\,GeV^{-1}\,\,\,{\left( \frac{m_a}{\mu \text{eV}} \right)}.\label{coupling mass}
\end{equation}
The above equation is a general one which is true for all kinds of axion model. Now,
making use of the relationship between axion coupling and mass, i.e. Eq. \eqref{mg}, as well as Eq. \eqref{coupling mass}, we provide the parameter space of axion particles in Fig.\ref{param1} for the result obtained from the CB effect and a specific model of sterile QCD axion. To plot this figure, we set the parameter $\beta=0.3^\circ$. In addition, it is sketched for two values of $\mathbb{F}=1$, assuming all kinds of CB effect come from the axion particles to get the threshold limit due to the CB on the axion's parameter space, and $\mathbb{F}=10^{-2}$. Based on this figure, the result of the CB effect cannot place any constraint on the allowed region of sterile QCD axion unless the axion's mass and the coupling constant are too small.

Before ending this part, it is noteworthy the fact that there will be an AC oscillation, at the axion
oscillation period 
\begin{equation}
	\tau_{osc} = \frac{2\pi}{m_a} \simeq 1.3\times 10^{-16} yr\,\,\,\,\,\,\,\,\,\,\,\,\,(\frac{\mu eV}{   m_a})
\end{equation}
 over the extended recombination epoch lasting $\tau_{rec} \sim 10^5$ yr, of the CMB polarization pattern on
the sky as measured today. Here, the sterile-neutrino QCD axion field was assumed to be a constant locally whose its vacuum expectation value is evaluated from the local DM energy density at a given redshift. However, on the assumption that ALPs play the role of DM, their vacuum expectation value oscillate over time, leading to induce time-dependent signals. This AC signal is quantitatively different from the CB induced by active and sterile neutrinos that discussed in the previous section.

\color{black}

\section{CB angle due to C$\nu$B and ALP}\label{combination}

\begin{figure}[t]
	\centering
	\includegraphics[scale=0.6]{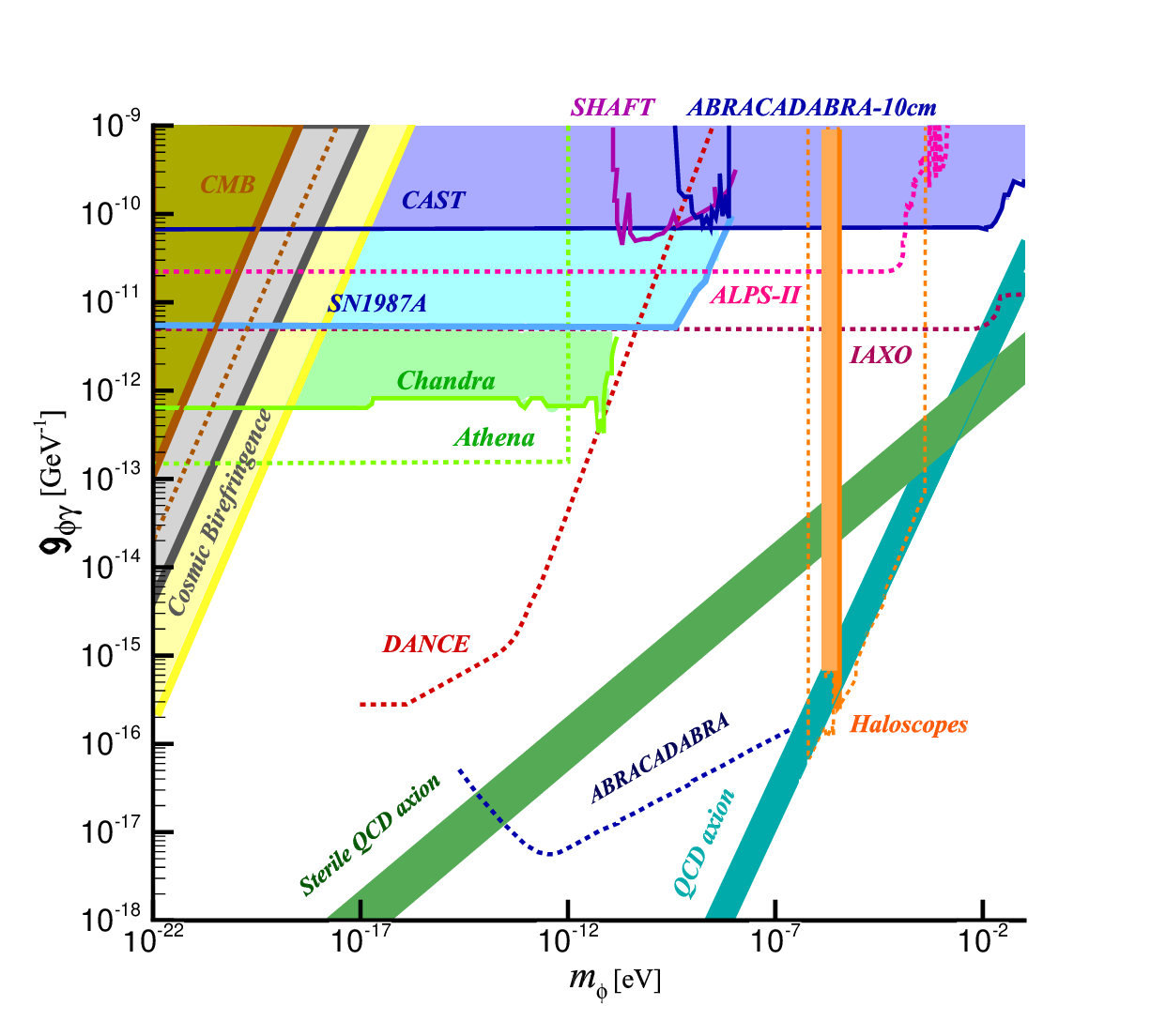}
	\caption{Parametric space of axion coupling ($\mathfrak{g}_{a\gamma}$)-mass ($m_a$). The figure is produced from fig.2 of Ref \cite{Michimura:2019qxr}. The yellow, light green and dark green areas show the uncertainty region due to the contribution of the cosmic neutrino interaction, Sterile QCD Axion and QCD Axion, respectively, in producing the CMB birefringence.} 
	\label{param}
\end{figure}

Here, we combine our previous results on CB angle calculations due to the interaction of the CMB photons with cosmic neutrinos as well as ALPs. Considering the fact that
\begin{equation}\label{beta1}
	\beta^{ob}=\beta_\nu+\beta_a, 
\end{equation}
where $\beta=0.342^{+0.094}_{-0.091}~\mathrm{deg}$ ($68\%\text{C.L.}$) indicates measured CB and employing the upper bound placed on the relative density asymmetry of neutrino and anti-neutrino ($\delta_\nu<3\times10^{-3}$), using the CB angle of the CMB, 
the axion CB contribution $\beta_a$ can be limited to consistent with $\beta^{ob}$. To this end, we choose $0.0001<\delta_\nu<0.0029$ and then we derive 
\begin{eqnarray}
	\label{min}&&\beta_a^{min}=\beta^{ob}-\beta_\nu(\delta_{\nu}=0.0029)=0.000248206~rad \label{betamin} \\
	&&\beta_a^{max}=\beta^{ob}-\beta_\nu(\delta_\nu=0.0001)=0.005058214~rad\label{max}.\label{betamax}
\end{eqnarray}
Now, extracting $\beta_{a}$ from Eq. \eqref{coupling mass} and using Eqs. \eqref{betamin} and \eqref{betamax}, the following results will be obtained for $\mathfrak{g}_{a\gamma}$ 
\begin{eqnarray}\label{max-min1}
	&&{\mathfrak{g}^{s}_{a\gamma}}_{min}\lesssim 0.312\times {10}\,GeV^{-1}{(\frac{m_a}{\mu eV})}	\nonumber\\
	&&{\mathfrak{g}^{s}_{a\gamma}}_{max}\lesssim 6.8\times \textcolor{blue}{10}\,GeV^{-1}{(\frac{m_a}{\mu eV})}	,
\end{eqnarray}
where $m_a$ is the mass of axion. These calculations show that the contribution of the ALPs -photon interaction in producing the CB angle can be faded by photon neutrino interaction. Indeed, due to the contribution of the C$\nu$B interactions in producing the CB effect, there will be an uncertainty on the parameter space of axions caused by this effect. The yellow area in Fig.{\ref{param}} shows the mentioned uncertainty region.

Before ending this section, it should be mentioned that we only considered the contribution of the C$\nu$B and ALPs interactions to discuss the constraint placed on the parameter space of axion. Some other mechanisms including polarized Compton scattering\cite{Khodagholizadeh:2019het}, dipole asymmetry of CMB temperature anisotropy \cite{Khodagholizadeh:2023sbv}, and various sorts of new physics interaction leading to produce CMB B mode polarization \cite{Khodagholizadeh:2023aft} that generates EB power spectrum, can contribute to the generation of the CB angle. However, by upgrading data on EB-cross power spectrum along with the theoretical and experimental studies in
the future we will be able to get more information about the origin of the reported CB angle. 


\section{Cross-power spectrum in the presence of Thomson scattering  and CMB-C$\nu$B and CMB-Axion}\label{cross}

\begin{figure}
	\centering
	{\includegraphics[scale=0.55]{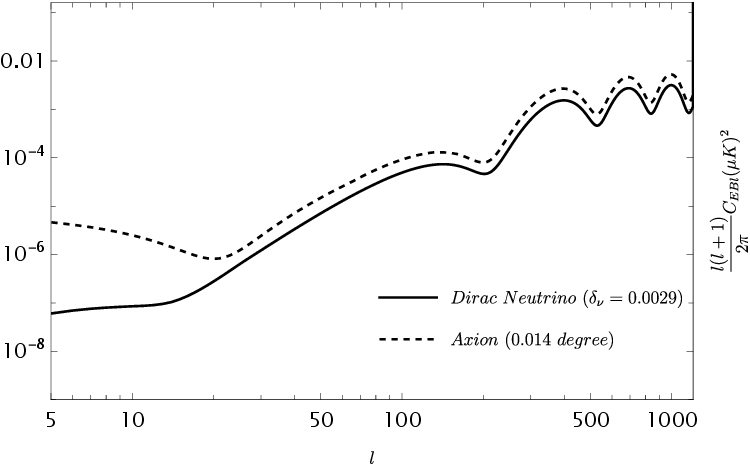}\label{fig3-1}} \hspace*{.1cm} 
	{\includegraphics[scale=0.5]{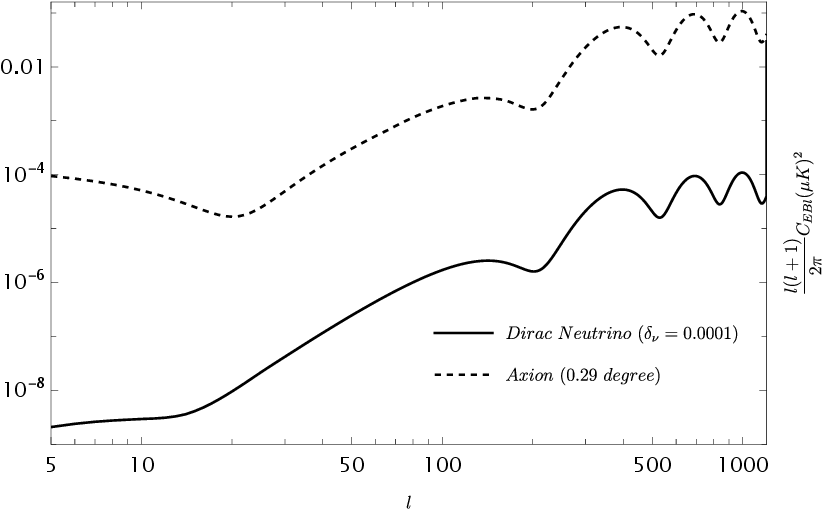}\label{fig3-2}}
	\caption{EB power spectrum via C$\nu$B-CMB and axion-CMB is plotted in terms of $l$ satisfying CB angle measurement constraint.}
	\label{EB}
\end{figure}

Since the CB effect violates the parity symmetry, it can generate the EB power spectrum of the CMB photons whose magnitudes, according to Eq. \eqref{ClEB}, depend on the integrated rotation of the polarization plane. Therefore, we should discuss, the contribution of the ALPs, as well as Dirac C$\nu$B, in producing the CB effect of the CMB through the EB cross-power spectrum. In this regard, we should solve the quantum Boltzmann equations \eqref{Boltzmann1} and \eqref{fps} for linear polarization of CMB photons in the presence of scalar metric perturbation. Solving these equations and making use of Eqs. \eqref{BE11} and \eqref{BE12}, the following results will be obtained: 
\begin{eqnarray}\label{delta}
	&&\Delta^{(\nu)}_E(K,\eta_0)=\int_0^{\eta_0}d\eta\,g(\eta)\Pi(k,\eta)\frac{j_\ell(x)}{x^2} \cos{\tau_\nu(\eta)},\\
	&&\Delta^{(\nu)}_B(K,\eta_0)=\int_0^{\eta_0}d\eta\,g(\eta)\Pi(k,\eta)\frac{j_\ell(x)}{x^2} \sin{\tau_\nu(\eta)},
\end{eqnarray}
\begin{eqnarray}
	&&\Delta_{E}^{(a)}(K, \eta_0) = \int_0^{\eta_0}
	d\eta g(\eta) \Pi(k,\eta) \frac{j_\ell(x)}{x^2}
	\cos{2\beta_{a}(\eta)},\\
	&&\Delta_{B}^{(a)}(K,\eta_0)=\int_0^{\eta_0}
	d\eta g(\eta) \Pi(k,\eta) \frac{j_\ell(x)}{x^2}
	\sin{2\beta_{a}(\eta)}.
\end{eqnarray}
where $g(\eta)=\dot{\tau}_{e\gamma}e^{-\tau_{e\gamma}}$ is the visibility function, describing the probability that a photon
scattered at epoch $\eta$ reaches the observer at the present time, $\eta_0$. Using the obtained solutions of the Boltzmann equation, we can calculate the combined power spectra among B-mode and E-mode as follows
\begin{eqnarray}
	C^{EB}_{ \ell(\nu,a)}&=&(4\pi)^2\frac{9}{16}\frac{(\ell+2)!}{(\ell-2)!}\int K^{2}dK P_S(K)\Delta^{(\nu,a)}_{E}\Delta^{(\nu,a)}_{B},\label{cross1}
\end{eqnarray}
By employing the above results, we have numerically calculated the EB cross power spectra for different values of $\tau_{\nu}$ and $\beta_{a}$, as are displayed in Fig. \ref{EB}. Note that the considered values for the parameters are such that the total contribution of C$\nu$B and ALPs interactions with photons in producing the CB angle of the CMB photons is consistent with the experimental measurement of the CB angle.

According to these figures, the behavior of the EB power spectrum of the CMB in the presence of C$\nu$B and ALPs interactions with photons are different in small $l$. Indeed, it is due to the fact that the redshift dependence of $\tau_{\nu}$ is such that the contribution of C$\nu$B interaction on the EB power spectrum will decrease, resulting in increasing the ALPs' contribution. Therefore, by increasing the sensitivity of the observational instrument and finding future observed data for $C_{l}^{EB}$, it becomes easier to distinguish between the values of the CB angle due to the different sources.

\section{Conclusion and Remarks} \label{conclusion}
In this paper, we studied some of probable sources that could contribute to producing the CB angle of the CMB. Our approach was based on solving the quantum Boltzmann equations in the presence of scalar perturbation of metric and extracting the beta angle in terms of the parameters of each theory.\par

First, we considered a sterile neutrino coupled to the SM gauge bosons $W^{\pm}$ and $Z^{0}$ through effective right-handed current interaction \eqref{Lagrangian} and supposed that this sort of sterile neutrino plays the role of DM in the Universe. Then, we calculated the photon-sterile neutrino forward scattering contribution to the CMB polarization. Calculations performed regarding the right-handed sterile neutrino DM revealed that this DM candidate could contribute to producing a part of the CB effect of the CMB photons. For instance, we concluded that a sterile neutrino DM with a mass of around $100KeV$ can contribute to generating CB up to $\beta\approx 1.5\times 10^{-9} rad$ at most. Next, we reviewed some results, presented in another paper \cite{Mohammadi:2021xoh} regarding the contribution of C$\nu$B-CMB interaction in producing the CB angle. The CB angle results comparison of sterile neutrino interaction shows very small contribution rather than the two other sources. \par

Moreover, we considered sterile neutrino QCD axion, which is a well-motivated solution to the strong CP problem, as a cold DM and investigated the effects of its interaction with photons on the polarization of the background photons. According to our calculation, this kind of interaction, if it exists, can rotate the polarization plane of the CMB and contribute to generating the CB angle. Afterwards, we used the reported CB angle of the CMB to put a new
constraint on the parameter space of sterile QCD axion (Fig. \eqref{param1}). \par

Next, we combined our results of CB angle calculations due to the interaction of the CMB photons with cosmic neutrinos as well as ALPs. Our calculations showed that the contribution of the ALPs-photon interaction in producing the CB angle can be faded by photon neutrino interaction. In fact, the contribution of the C$\nu$B interactions in producing the CB effect leads to an uncertainty on the parameter space of axions. For more clarification, we provided Fig.{\ref{param}}, in which the full mass dependence of the result has been illustrated.
Note that this figure is originally adapted from Ref. \cite{Michimura:2019qxr}, and we added our result to it. In addition, we discussed the contribution of the ALPs as well as Dirac C$\nu$B in producing the CB effect of the CMB through the EB cross-power spectrum. The outcome was that the EB power spectrum of the CMB behaves differently in the presence of C$\nu$B and ALPs interactions in small $l$. Hence, we hope that by increasing the sensitivity of the observational instrument and finding future observed data for $C_{l}^{EB}$, we will be able to make a difference between the value of the CB angle due to the various sources of its production.

\appendix
\section{Appendix}\label{appendix}

Sterile neutrino couplings (\ref{Lagrangian}) to SM gauge bosons are right-handed, which are distinct from active neutrino left-handed couplings. Therefore, we need to do calculations analogously to the case of active neutrinos \cite{Khodagholizadeh:2023aft}. In this appendix, we present detailed calculations of sterile neutrino contributions to the CB angle.

\subsection{Calculation of Time evolution of the density matrix components via photon-Sterile neutrino interaction}

\begin{figure}[tb]
	\center
	\includegraphics[scale=1]{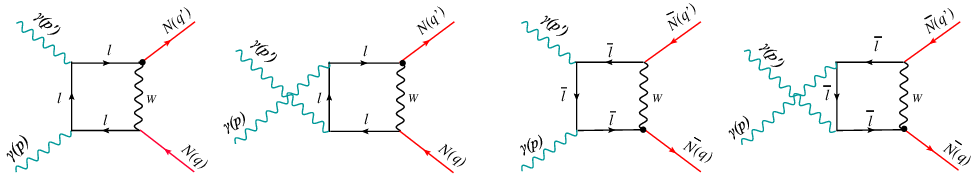}\\
	\caption{The representative Feynman diagrams represent the photon-sterile neutrino scattering. In the first SM family, $\ell=e$ and $N^\ell=N^e=N$, $\bar{\ell}$ and $\bar{N}$ indicate their anti-particles. The sterile neutrino mass $M_{\text{\tiny{N}}}=M_{\text{\tiny{N}}^e}$.} \label{feyn}
\end{figure}

Based on the effective Lagrangian introduced in \eqref{Lagrangian}, the photon can scatter from sterile neutrino particles. Indeed, the leading order  
$\mathcal O({\mathcal{G}}^2_R)$ contribution comes from the scattering of photons from sterile neutrinos at a one-loop level with a lepton and weak gauge bosons propagating in the loop. Representative relevant Feynman diagrams are shown in Fig. \ref{feyn}.\par

To calculate the time evolution of the density matrix components due to the forward scattering of the photon-sterile neutrino interaction, we need the
Fourier transformations of the electromagnetic free gauge field $A^{\mu}$ and Majorana fermion field $N(x)$, having self-conjugate property, which are given as follows
\begin{equation}\label{gauge}
	A_{\mu}(x)=\int\frac{d^3 {\bf k}}{(2\pi)^3 2k^0}[\mathfrak{a}_{s}(p)\epsilon_{s\mu}(k)e^{-ik.x} + \mathfrak{a}_{s}^\dagger (k)\epsilon_{s\mu}^* (k)e^{ik.x}],
\end{equation}
\begin{equation}\label{majorana1}
	N(x)= \int \frac{d^3{\bf q} }{(2\pi)^3}\frac{M_{\tiny{\text{N}}}}{ q^0}\left[ \mathfrak{b}_r(q) \mathcal{U}_{r}(q)
	e^{-iq\cdot x}+ \mathfrak{b}_r^\dagger (q) \mathcal{V}_{r}(q)e^{iq\cdot x}
	\right],
\end{equation}
where $\epsilon_{s\mu}(p)$  with $s=1,2$ are the photon polarization 4-vectors of two physical transverse polarization while $\mathcal{U}_{r}(q)$ and $\mathcal{V}_{r}(q)$ are the Dirac spinors. Moreover, the creation $\mathfrak{a}_{s}^\dagger(k)$ ($\mathfrak{b}_{r}^\dagger(q)$) and annihilation $\mathfrak{a}_{s}(k)$ ($\mathfrak{b}_{r}(q)$) operators obey the following canonical commutation (anti-commutation) relations
\begin{eqnarray}
	[\mathfrak{a}_{s}(k),\mathfrak{a}_{s'}^\dagger (k')]&=&(2\pi)^3 2k^0 \delta_{ss'}\delta^{(3)}({\bf k}-{\bf k'}),\nonumber \\
	\{\mathfrak{b}_{r}(q),\mathfrak{b}_{r'}^\dagger (q')\} &=&(2\pi)^3 \frac{q^0}{M_{\tiny{\text{N}}}} \delta_{rr'} \delta^{(3)}({\bf q}-{\bf q'}).
\end{eqnarray}
The leading-order interacting Hamiltonian for the scattering, represented in Fig. \ref{feyn}, is found by making use of \eqref{HI}, \eqref{Lagrangian}, and the above relations, which is given by the following scattering amplitude
\begin{eqnarray}\label{H1}
	H_{I}^0(t)&=&\,\int {d\bf{q}} {d\bf{q'}} {d\bf{k}} {d\bf{k'}}(2\pi)^3 \delta^{(3)} ({\bf{q'}}+{\bf{k'}}-{\bf{q}}-{\bf{k}}) \exp({i[q'^0 +k'^0 -q^0 - k^0]})\nonumber\\
	&\times&[\mathfrak{b}_{r'}^\dagger({\bf q'}) \mathfrak{a}_{s'}^\dagger({\bf k'})\mathcal{M}_\text{tot}(N\gamma\,\to\,N\gamma)\, \mathfrak{a}_{s}({\bf k}) \mathfrak{b}_{r}({\bf q})],
\end{eqnarray}
with ${d \bf{q}} \equiv \frac {d^3 \bf{q}}{(2\pi)^3}\frac{M_{\tiny{\text{N}}}}{q^0}$, ${d \bf{k}} \equiv \frac {d^3 \bf{k}}{(2\pi)^3}\frac{1}{2 k^0}$ and the total amplitude ${M}_\text{tot}$ can be obtained from the sum of all Feynman diagrams in Fig.\ref{feyn}, as follows 
\begin{eqnarray}\label{m}
	{M}_{tot}({\bf q'}r',{\bf k'}s',{\bf q}r,{\bf k}s)&\equiv &{M}_{1}({\bf q'}r',{\bf k'}s',{\bf q}r,{\bf k}s) +{M}_{2}({\bf q'}r',{\bf k'}s',{\bf q}r,{\bf k}s)\nonumber\\
	&&-{M}_{3}({\bf q'}r',{\bf k'}s',{\bf q}r,{\bf k}s) -{M}_{4}({\bf q'}r',{\bf k'}s',{\bf q}r,{\bf k}s),
\end{eqnarray}
where ${M}_{3,4}({\bf q'}r',{\bf k'}s',{\bf q}r,{\bf k}s)$ are, respectively,  the Hermitian conjugates of ${M}_{1,2}({\bf q'}r',{\bf k'}s',{\bf q}r,{\bf k}s)$ and have been contributed from  antiparticles in the loops as follows 
\begin{eqnarray}\label{m1}
	{M}_{1}({\bf q'}r',{\bf k'}s',{\bf q}r,{\bf k}s)&=&\frac{1}{(2\pi)^{4}}\,\frac{e^{2}\,g_{\tiny\text{W}}^{2}}{8}\,\mathcal{G}_{R}^{2}\int d^{4}l\,\, \,\bar{\mathcal{U}}_{r^{'}}({\bf q^{'}})\gamma^{\alpha}\,(1-\gamma^{5})\,\mathcal{S}_{F}(l+k-k^{'})\slashed{\epsilon}_{s^{'}}({\bf k^{'}})\nonumber\\
	&&\mathcal{S}_{F}(k+l)\,\slashed{\epsilon}_{s}({\bf k})\,\mathcal{S}_{F}(l)\gamma ^{\beta}\,(1-\gamma^{5})\,\mathcal{U}_{r}({\bf q})\,D_{F_{\alpha\beta}}(q-l),
\end{eqnarray}
\begin{eqnarray}\label{m2}
	{M}_{2}({\bf q'}r',{\bf k'}s',{\bf q}r,{\bf k}s)&=&\frac{1}{(2\pi)^{4}}\,\frac{e^{2}\,g_{\tiny\text{W}}^{2}}{8}\,\mathcal{G}_{R}^{2}\int d^{4}l\,\, \,\bar{\mathcal{U}}_{r^{'}}({\bf q^{'}})\gamma^{\alpha}\,(1-\gamma^{5})\,\mathcal{S}_{F}(l+k-k^{'})\slashed{\epsilon}_{s}({\bf k})\nonumber\\&&\mathcal{S}_{F}(l-k^{'})\,\slashed{\epsilon}_{s^{'}}({\bf k^{'}})\,\mathcal{S}_{F}(l)
	\gamma ^{\beta}\,(1-\gamma^{5})\,\mathcal{U}_{r}({\bf q})\,D_{F_{\alpha\beta}}(q-l),
\end{eqnarray}
\begin{eqnarray}\label{m3}
	{M}_{3}({\bf q'}r',{\bf k'}s',{\bf q}r,{\bf k}s)&=&\frac{1}{(2\pi)^{4}}\,\frac{e^{2}\,g_{\tiny\text{W}}^{2}}{8}\,\mathcal{G}_{R}^{2}\int d^{4}l\,\, \,\bar{\mathcal{V}}_{r}({\bf q})\gamma^{\alpha}\,(1+\gamma^{5})\,\mathcal{S}_{F}(-l)\slashed{\epsilon}_{s}({\bf k})\mathcal{S}_{F}(-k-l)\nonumber\\
	&&\,\slashed{\epsilon}_{s^{'}}({\bf k^{'}})\,\mathcal{S}_{F}(k^{'}-k-l)\gamma ^{\beta}\,(1+\gamma^{5})\,\mathcal{V}_{r}({\bf q'})\,D_{F_{\alpha\beta}}(l-q),
\end{eqnarray}
and 
\begin{eqnarray}\label{m4}
	{M}_{4}({\bf q'}r',{\bf k'}s',{\bf q}r,{\bf k}s)&=&\frac{1}{(2\pi)^{4}}\,\frac{e^{2}\,g_{\tiny\text{W}}^{2}}{8}\,\mathcal{G}_{R}^{2}\int d^{4}l\,\, \,\bar{\mathcal{V}}_{r}({\bf q})\gamma^{\alpha}\,(1+\gamma^{5})\,\mathcal{S}_{F}(-l)\slashed{\epsilon}_{s^{'}}({\bf k^{'}})
	S_{F}(k^{'}-l)\nonumber\\
	&&\,\slashed{\epsilon}_{s}({\bf k})\,\mathcal{S}_{F}(k^{'}-k-l)\gamma ^{\beta}\,(1+\gamma^{5})\,\mathcal{V}_{r^{'}}({\bf q^{'}})\,D_{F_{\alpha\beta}}(l-q),
\end{eqnarray}
where $\mathcal{S}_{F}$ denotes the fermion propagator, and the indices $r,r'$ and $s,s'$ stand for the sterile neutrino and photon spin states, respectively.
Now, in order to calculate the forward scattering term in (\ref{forward}), we need to find the commutator $[H_I^0(t),D_{ij}^0({\bf p})]$, then evaluate the expectation value $\langle[H_I^0(t),D_{ij}^0({\bf p})]\rangle$  according to the following operator expectation value
\begin{equation}
	\langle \, \mathfrak{b}^\dag_{r'_{i}}(q')\mathfrak{b}_{r_{j}}(q)\, \rangle
	=(2\pi)^3\delta^3(\mathbf{q}-\mathbf{q'})\delta_{rr'}\delta_{ij}\frac{1}{2}f_\text{\tiny{N}}(\mathbf{x},\mathbf{q}).
\end{equation}
To this end, one can substitute (\ref{m}-\ref{m4}) into (\ref{H1}) and then (\ref{forward}) to find the time evolution of the density matrix components which is obtained as follows
\begin{eqnarray}\label{rhodott}
	\frac{d}{dt}\rho_{ij}(k)\,&=&\,-\frac{\sqrt{2}}{12\,\pi\,k^{0}}\alpha\,\mathcal{G}_{R}^{2}\,{G}_{\text{\tiny{F}}}\int d{\bf{q}}\,\, (\delta_{is}\rho_{s'j}(k)-\delta_{js'}\rho_{is}(k))\,f_\text{\tiny{N}}({\bf{x}},{\bf{q}})\,
	\bar{\mathcal{U}}_{r}(q)\,\,(1-\gamma^{5})\nonumber\\&&(q\cdot\epsilon_{s}\,\,\slashed{\epsilon}_{s^{'}}\,+\,q\cdot\epsilon_{s^{'}}\,\,\slashed{\epsilon}_{s})\,\mathcal{U}_{r}(q)+\frac{\sqrt{2}}{24\,\pi\,k^{0}}\alpha\,\mathcal{G}_{R}^{2}\,{G}_{\text{\tiny{F}}}\int d{\bf{q}}\,\, (\delta_{is}\rho_{s'j}(k)-\delta_{js'}\rho_{is}(k))\,\nonumber\\&&f_\text{\tiny{N}}({\bf{x}},{\bf{q}})\,    \bar{\mathcal{U}}_{r}(q)\,(1-\gamma^{5})\,\slashed{k}\,(\slashed{\epsilon}_{s^{'}}\,\slashed{\epsilon}_{s}\,-\,\slashed{\epsilon}_{s}\,\slashed{\epsilon_{s^{'}}})\,\mathcal{U}_{r}(q).
\end{eqnarray}
where $f_\text{\tiny{N}}({\bf{x}},{\bf{q}})$ is the distribution function of Strile neutrinos. Now by using of this equation one can compute the time deviation of the Stokes parameters and polarization vector $\dot\Delta _{P}^{(N)}$.

\subsection{Sterile neutrinos' impact on the CB angle}

With the results of \cite{Khodagholizadeh:2023aft}, Stokes parameters can be constructed using \eqref{eq:rho}, showing that this interaction can affect the evolution of the linear polarization of the CMB as follows:
\begin{eqnarray}
	\dot\Delta _{P}^{(N)} +iK\mu \Delta _{P}^{(N)} 
	= \dot\tau_{e\gamma}\Biggl[ -\Delta^{(N)}_{P}
	\left .\sum_{m}\sqrt{\frac{6\pi}{5}}
	\ _{\pm 2} Y_2^m
	S_{P}^{(m)} \right] \mp a(\eta) \dot{\tau}_N(\eta) \Delta _{P}^{\pm (N)},
	\label{Boltzmann1}
\end{eqnarray}
where superscripts $N$ stands for sterile neutrino, $\Delta _{P}^{(N)}=Q^{(N)}\pm i U^{(N)}$, $S_P^{(m)}({\bf
	k},\eta) \equiv \Delta_{T2}^{(m)}({\bf k},\eta)+
12\sqrt{6}\Delta_{+,2}^{(m)}({\bf k},\eta)+12\sqrt{6}\Delta_{-,2}^
{(m)}({\bf k},\eta)$ is the source term of generating polarization. The time variation of the effective neutrino opacity ${\tau}_\text{\tiny{N}}$ is 
\begin{eqnarray}\label{eta}
	\dot{\tau}_\text{\tiny{N}}&=&\frac{\sqrt{2}}{3\pi k^{0}\,M_{\text{\tiny{N}}}}\,\,\alpha\,\,{G}_{\text{\tiny{F}}}\,\mathcal{G}_{R}^{2}\,\int\,d{\bf{q}}\,f_\text{\tiny{N}}({\bf x},{\bf q})\,\times (\varepsilon _{\mu\,\nu\,\rho\,\sigma}\epsilon_{2}^{\mu}\,\epsilon_{1}^{\nu}\,k^{\rho}\,q^{\sigma}),
\end{eqnarray}
where it can be reduced to 
\begin{eqnarray}
	\dot{\tau}_\text{\tiny{N}}&=&  \frac{\sqrt{2}}{3\pi k^{0}\,M_{\text{\tiny{N}}}}\,\,\alpha\,\,{G}_{\text{\tiny{F}}}\,\mathcal{G}_{R}^{2}\int d\mathbf{q}\, f_\text{\tiny{N}}({\bf x},{\bf q})
	\times \left[q^0 {\bf k}\cdot(\epsilon_1\times\epsilon_2)+k^0 {\bf q}\cdot(\epsilon_1\times\epsilon_2)\right]\nonumber\\
	&=&  \frac{\sqrt{2}}{3\pi}\,\,\alpha\,\,{G}_{\text{\tiny{F}}}\,\mathcal{G}_{R}^{2}\,n_{\text{\tiny{N}}}\left[1+\langle {\bf v}\rangle\cdot(\epsilon_1\times\epsilon_2)\right]\approx \frac{\sqrt{2}}{3\pi}\,\,\alpha\,\,{G}_{\text{\tiny{F}}}\,\mathcal{G}_{R}^{2}\,n_{\text{\tiny{N}}},\label{Bmode1}
\end{eqnarray}
where ${\bf k}\cdot(\epsilon_1\times\epsilon_2)= |{\bf k}|$, the sterile neutrino number density $n_\text{\tiny{N}}=\int \frac{d^3{\bf{q}}}{(2\pi)^3} f_\text{\tiny{N}}({\bf x},{\bf q})$, and $ \langle {\bf v}\rangle $ is the average velocity of sterile neutrino particles. Henceforth, we suppose that this sort of sterile neutrino plays the role of DM in the Universe. Since the average velocity of DM is small, the dominant contribution of
this scattering to photon polarization comes
from the first term and thus we ignore the term including $ \langle {\bf v}\rangle $. \par

Using the fact that
\begin{equation}\label{BE6}
	{\tau}_{\text{\tiny{N}}}(\eta,\mu)\equiv\int_{0}^{\eta}d \eta a(\eta)\dot{\tau}_{\text{\tiny{N}}},
\end{equation}
along with the following cosmological relations
\begin{equation}\label{redshift}
	a\,d\eta=-\frac{dz}{H(z)(1+z)},
\end{equation}
and 
\begin{equation}\label{Hubble}
	\frac{H^2}{H_{0}^2}=\Omega_{M}^0(1+z)^3+\Omega_{\Lambda}^0,
\end{equation}
in which $H(z)$ is the Hubble parameter and $z$ is the cosmological redshift, we arrive at the following equation for the effective sterile neutrino opacity 
\begin{equation}
	\tau_{\text{\tiny{N}}}=\frac{\sqrt{2}}{3\pi\,M_{\text{\tiny{N}}}}\,\,\alpha\,\,{G}_{\text{\tiny{F}}}\,\mathcal{G}_{R}^{2}\,\rho_{\text{\tiny{N}}}^{0}\,\frac{2H(z')}{3\Omega_M^0H_0^{2}}\Big|^{z'=z}_{z'=0},
	\label{tauN}
\end{equation}
\color{black}
where we have used the fact that $\rho_{\text{\tiny{N}}}\,=\,\rho _{\text{\tiny{N}}}^0 (1+z)^3$ with $\rho _{\text{\tiny{N}}}^0$ being the mass density of DM in the present time. Considering the point that $H_{0}\,=\, (67.4\,\pm\,0.5)\, \text{km}\text{s}^{-1}\text{Mpc}^{-1}$,\,\,$\Omega_{M}^0\,=\,0.315\,\pm\,0.007$, \,\,$\Omega_{\Lambda}^0\approx0.69$ \cite{Planck:2018vyg}, we estimate the maximum value $\tau_{\text{\tiny{N}}}$ near the last scattering, which leads to eq.~(\ref{tauN1}) in the main text.

\end{document}